\documentclass[twocolumn,showpacs,preprintnumbers,amsmath,amssymb]{revtex4}

\usepackage{graphicx}
\usepackage{dcolumn}
\usepackage{bm}

\begin{document}


\title{SU(m) non-Abelian anyons in the Jain hierarchy of quantum Hall states}
\author{M. Cristina Diamantini}
\email{cristina.diamantini@pg.infn.it}
\affiliation{%
INFN and Dipartimento di Fisica, University of Perugia, via A. Pascoli, I-06100 Perugia, Italy
}%


\author{Carlo A. Trugenberger}
\email{ca.trugenberger@bluewin.ch}
\affiliation{%
 SwissScientific, ch. Diodati 10 CH-1223 Cologny, Switzerland
}%


\date{\today}

\begin{abstract}
We show that different classes of topological order can be distinguished by the dynamical symmetry algebra of edge excitations. Fundamental topological order is realized when this algebra is the largest possible, the algebra of quantum area-preserving diffeomorphisms, called $W_{1+\infty}$. We argue that this order is realized in the Jain hierarchy of fractional quantum Hall states and show that it is more robust than the standard Abelian Chern-Simons order since it has a lower entanglement entropy due to the non-Abelian character of the quasi-particle anyon excitations. These behave as SU($m$) quarks, where $m$ is the number of components in the hierarchy. We propose the topological entanglement entropy as the experimental measure to detect the existence of these quantum Hall quarks. Non-Abelian anyons in the $\nu = 2/5$ fractional quantum Hall states could be the primary candidates to realize qbits for topological quantum computation. 

\end{abstract}

\maketitle

The quantum Hall effect \cite{lau} is one the of most striking examples of emergence in physics. A large number of electrons conspire to organize themselves into macroscopic quantum states which manifests themselves in a series of rational filling fractions $\nu$ in the Hall conductances $\sigma_H=\nu e^2/h$. These are measured to an astonishing precision in terms of the standard unit $e^2/h$. 

Universality and precision in macroscopic states are normally associated either with topological protection or with a large dynamical symmetry group. Indeed, both these reasons have been proposed to explain the observed emergent behavior of quantum Hall fluids.

On one side it has been proposed that quantum Hall fluids realize a new type of topological order \cite{wen} describing a particular entanglement of the electrons characterized by a gap for all bulk excitations, by a finite degeneracy on topologically non trivial spaces and by the appearance of quasi-particle excitations with fractional charge and statistics. 

On the other side, the quantum incompressibility of Hall fluids has been related to the presence of an infinite-dimensional dynamical symmetry under the algebra $W_{1+\infty}$ of quantum area-preserving diffeomorphisms \cite{ctz1}. The observed universality and precision stem from the infinite (in the thermodynamic limit) number of highest-weight conditions on the ground state \cite{ctz1,ctz2} and the dynamical symmetry completely determines the spectrum of  excitations of the theory \cite{ctz2}.

The two approaches agree on the spectrum of quantum numbers of excitations. The $W_{1+\infty}$ dynamical algebra, however predicts non-Abelian anyons of a new type with respect to \cite{wen3}, an issue which is far from academic due to its possible practical relevance for  topological quantum computation \cite{kitaev}. 

In this paper we clarify the relation between topological order and dynamical symmetry of the edge excitations. We argue that there are different classes of topological order, characterized by the corresponding dynamical symmetry algebra of edge excitations. At the most fundamental level, topological order is equivalent to $W_{1+\infty}$ dynamical symmetry of edge excitations. 
This  generalizes the connection between bulk Chern-Simons theories and edge conformal field theories \cite{wen1} to  new topological field theories and maximally extended conformal algebras. In other classes of topological order, the edge dynamical symmetry can be broken by quantum effects to a subgroup of $W_{1+\infty}$.

Secondly, we continue the analysis of the $W_{1+\infty}$ minimal models \cite{ctz2} by computing the quantum dimensions of their non-Abelian anyon excitations. The physical picture that emerges from this computation is that of "quark-like" excitations carrying an SU($m$) isospin quantum number, where m is the number of edges in the hierarchical fluid \cite{jain,wen}. 

Finally, we compute the entanglement entropy \cite{sre} of the minimal models and show that it is lower than the corresponding quantity computed from Chern-Simons topological order due to the presence of non-Abelian excitations. This is the formalization of the larger stability of the ground states corresponding to the $W_{1+\infty}$ minimal models with respect to their Chern-Simons topological order counterparts. 

Topological order arises in many-body systems whose long-distance effective field theory is trivially invariant under diffeomorphisms since it does not depend on the metric of space-time, the main  examples being Chern-Simons models in 2+1 dimensions and BF theories \cite{birmi}  in any number of dimensions. It has been recently shown that topologically ordered media in 2+1 dimensions are characterized by an entaglement entropy \cite{sre}
\begin{equation}S = aL-\gamma \ ,
\label{a}
\end{equation}
where $L$ is the length of the one-dimensional boundary $\delta D$ of the disk-like region $D$ on which the state is defined and $\gamma$ is a universal term called the topological entanglement entropy. 

The entanglement entropy is a characteristics of the quantum ground state at energies well below the gap and is thus preserved by the gapless fluctuations corresponding to edge excitations living on $\delta D$.  The largest group of symmetry transformations of the world sheet $R \times \delta D$ that preserves the length $L$ of $\delta D$ at fixed time and contains the two-dimensional Poincar\'e group is that of classical area-preserving diffeomorphisms, whose algebra is called $w_{\infty}$ \cite{pope}. Correspondingly, the largest symmetry algebra of the (1+1) dimensional quantum field theory of the edge excitations of a topologically ordered medium with a given, fixed entanglement entropy is the quantum version $W_{1+\infty}$ (or  $W_{1+\infty} \times \bar W_{1+\infty}$ in the non-chiral case) of the algebra of area-preserving diffeomorphisms \cite{pope}. 

$W_{1+\infty}$ contains the Virasoro algebra of conformal transformations as a subalgebra. Actually, $W_{1+\infty}$ is the maximal extension of the conformal algebra, obtained by adding infinite currents of spin $s$ for all $s =3...\infty$ to the Virasoro algebra for the spin 2 current (the "1" in the notation $W_{1+\infty}$ corresponds to the inclusion of the U(1)  current of spin 1) \cite{scho}. 

We now identify as fundamental topological order all entanglement patterns that are characterized by the full $W_{1+\infty}$ dynamical symmetry of edge excitations. Other classes of topological order correspond to situations in which the $W_{1+\infty}$ symmetry is broken to a subgroup. The smaller this subgroup, the less robust is the corresponding topological order. 

It has been shown that all observed fractional quantum Hall fluids of the Jain hierarchy \cite{jain} have fundamental topological order in this sense. Indeed, all the chiral boson theories of the edge excitations corresponding to the Abelian Chern-Simons models describing the bulk long-distance field theories for these states possess the entire $W_{1+\infty}$ dynamical symmetry. This is true also for the observed even-denominator $\nu = 5/2$ paired Haldane-Rezayi state \cite{cgt}. It is not true, instead, for the Pfaffian state corresponding to the $\nu =1/2$ state. In this case, the $W_{1+\infty}$ symmetry is still present at the semiclassical level, but is broken down by quantum effects to a $c=3/2$ conformal field theory \cite{cgt}. In this sense we would assign the $\nu =1/2$ state to a different, less robust class of topological order. It is indeed well known, that the $\nu=1/2$ state has a different character than all other observed fractional quantum Hall fluids and corresponds more to a standard Fermi sea of composite fermions than to an incompressible quantum fluid \cite{lau}. 

In the standard approach to topological order in fractional quantum Hall fluids, the long distance effective field theories describing the composite fermion hierarchy of Jain at filling fraction $\nu = m/mp\pm 1$, $p$ even, are taken as linear combinations of $m$ interacting Chern-Simons fields describing the $m$ components of the fluid. The corresponding edge excitations are described by $m$ interacting chiral bosons that form  affine $\widehat U(1)^m$ representations and can be reorganized into $\widehat U(1) \times \widehat {SU}(m)_1$ representations (which are always Abelian).  

From the point of view of the conformal algebra, these representations are always fundamental. From the point of view of the extended $W_{1+\infty}$ algebra, however, there are cases in which these representations are degenerate \cite{kac}. In these cases the $\widehat U(1)^m$ representations can be reduced to representations of the $c = m$ $\widehat U(1) \times W_m(p\to \infty)$ minimal models by projecting out all null vectors. Here, $W_m$ is the Fateev-Lykyanov-Zamolodchikov algebra \cite{scho}. 

The fusion rules of $W_m$ representations are isomorphic to the decomposition of SU($m$) tensor representations. By assembling minimal sets of degenerate representations which are closed under these fusion rules one constructs the $W_{1+\infty}$ minimal models \cite{ctz2}. Each such minimal model describes a self-consistent set of edge excitations which are all related by bootstrap and which have full $W_{1+\infty}$ dynamical symmetry: these sets of excitations are called "minimal" since they represent the smallest possible number of edge excitations consistent with symmetry under quantum area- preserving diffeomorphisms. In particular they contain less states than the generic $W_{1+\infty}$ models, which are equivalent to the standard Chern-Simons-based models. A crucial consequence of this reduction of degenerate $\widehat U(1)^m$ representations to $\widehat U(1)\times W_m$ ones is that the statistics of excitations becomes non-Abelian. Each $W_{1+\infty}$ minimal model defines a maximal type of fundamental topological order. The crucial and suggestive fact is that the $W_{1+\infty}$ minimal models exists only and exactly for the filling fractions corresponding to the Jain hierarchy of the fractional quantum Hall effect \cite{ctz2}.

The relation between the Abelian Chern-Simons models and the $W_{1+\infty}$ minimal models can be summarized in the spectrum of the corresponding edge excitations. 
Both models are defined for filling fractions $\nu = m/mp\pm1$, $p>0$ even, and have central charge $c=m$ corresponding to the integer number of components. Both models predict excitations with charges and fractional statistics parameters given by:
\begin{eqnarray}
Q &=& {1\over mp\pm 1} \sum_{i=1}^m n_i \ ,
\nonumber \\
{\theta \over \pi} &=& \pm \left( \sum_{i=1}^m n_i^2 - {p\over mp\pm 1} \left( \sum_{i=1}^m n_i \right)^2 \right) \  ,
\label{b}
\end{eqnarray}
which agree with experimental data and match the results of the Jain hierarchy \cite{jain}. 

The differences between the models are three \cite{ctz2}. The first concerns the span of the lattice of excitations: this is unrestricted, $n_i \in Z$, for the Abelian Chern-Simons model while the condition $n_1 \ge n_2 \ge \dots \ge n_m$ holds for the  $W_{1+\infty}$ minimal models. The second difference lies in the fusion rules of excitations. These are Abelian for the Chern-Simons model; denoting by $\{n\}$ the tuple $(n_1,\dots , n_m)$ we have $\{n\} \bullet \{k\} = \{n+k\}$. In the $W_{1+\infty}$ minimal models the fusion rules are non-Abelian. Let us denote by $q(n)=n_1+\dots +n_m$ and by $\Lambda = \sum_{a=1}^{m-1} \Lambda^{(a)} (n_a-n_{a+1})$ , where $ \Lambda^{(a)}$ are the fundamental weights of SU(m). Then $q(n)$ is simply summed under fusion exactly as in the Abelian Chern-Simons models, while $\Lambda$ behaves under fusion as the highest weight vector of an SU($m$) representation. Finally, in the Abelian Chern-Simons model there is no a priori reason to choose the filling fractions corresponding to the Jain hierarchy. These are instead predicted as corresponding to the most stable topological order in the $W_{1+\infty}$ minimal models. 

One puzzling question concerning the physical content of the $W_{1+\infty}$ minimal models is the apparent absence of the states needed to form full $SU(m)$ multiplets. At first sight it looks like every representation in the $W_{1+\infty}$ minimal models contains only the highest weight state of an SU($m$) multiplet but not the other states.

Here we show that this is not the case. The states needed to form complete SU($m$) multiplets are indeed present: they are simply hidden among the infinite tower of states in the Verma modules of the infinite-dimensional algebra $W_{1+\infty}$. 

To show this we compute the quantum dimensions of degenerate $W_{1+\infty}$ representations. When dealing with infinite-dimensional algebras the concept of dimension of a representation does not make any sense, since all representations contain an infinite number of states. The best alternative is to regularize the number of states in a representation by subtracting the infinite number of states in the ground state. This defines the quantum dimension $\cal D$ of the representation: $\cal D$ measures the excess of states in a representation with respect to the ground state and is defined as \cite{dms}:
\begin{equation}
{\cal D}_M = {\rm lim}_{q \to 1}{\chi_{M (W_{1+\infty}, m, \vec{ r})} (q) \over \chi_{M_0 (W_{1+\infty}, m, \vec{ 0})}(q)} \ ,
\label{c}
\end{equation}
where $\chi_{M (W_{1+\infty}, m, \vec r)} (q)$ is the character of the $W_{1+\infty}$
representation $M$ with central charge $m$ and weight vector $\vec r = (r_1 \dots r_m)$.

The characters of degenerate $W_{1+\infty}$ representations are known \cite{kac,ctz2}:
\begin{eqnarray}
&\chi_{M (W_{1+\infty}, m, \vec r)}& (q) = {q^{\sum_{i=1}^m r_i^2/2}\over \eta(q)^m}
\prod_{1\le i <j \le m} \left( 1-q^{n_i-n_j+j-1} \right) \ , 
\nonumber \\
&\vec r = (r_1, \dots, r_m)& = (s+n_1, \dots, s+n_m),  s \in R \ ,
\label{d}
\end{eqnarray}
where $\eta (q)$ is the Dedekind function, $s$ is an unimportant real parameter and the $n_i$ are the integers defining the representation as explained above. 

After some algebra it is possible to rewrite (\ref{c}), in the limit $q \to1$, as
\begin{equation}
{\cal D}_M = {\prod_{ k = 1}^{m - 1}  \prod_{ i = 1}^{m - 1} \left( k + n_i - n_{i + k} \right) \over  \prod_{ k = 1}^{m - 1}  k^{m - k}} \ .
\label{qdw}
\end{equation}
We can now compare this to the classical dimensions $D$ of $SU(m)$ representations. To this end we recall that these  can be written as \cite{wyb}:
\begin{equation}
D_{SU(m)} =  \prod_{1 \le q < p \le m} { r_p - r_q \over g_q - g_p}  \ ,
\label{qds}
\end{equation}
where: $r_i = l_i + g_i$ and $g_i = (m-1)/ 2 - i + 1$ and the integers $l_i$ can be chosen such that $l_1 \ge l_2 \ge \dots \ge l_n \ge 0$ .   
Noticing that $g_q - g_p = p -q$ we rewrite (\ref{qds}) as:
\begin{equation}
D_{SU(m)} =  \prod_{1 \le q < p \le m} {p-q + r_q -r_p \over p- q}  \ .
\label{nqds}
\end{equation}
Setting now $q = i$ and $p = i + k$ we can rewrite ( \ref{nqds}), after some algebra as:
\begin{equation}
D_{SU(m)} =  {\prod_{ k = 1}^{m - 1}  \prod_{ i = 1}^{m - 1} \left( k + r_i - r_{i + k} \right) \over  \prod_{ k = 1}^{m - 1}  k^{m - k}} \ ,
\label{qdu}
\end{equation}
from which we see that the quantum dimensions ${\cal D}_M$ coincide with the classical ones of $SU(m)$,  $D_{SU(m)}$, with the same highest weight $\Lambda$.

This computation shows that the excess states in a degenerate $W_{1+\infty}$ representation with respect to the ground state form the full SU($m$) multiplet corresponding to the highest weight vector $\Lambda$ defining the representation. It is almost inevitable to interpret the infinity of states common to all representations as the gapless edge excitations and the SU($m$) multiplets as the edge manifestation (due to incompressibility) of bulk quasi-particles. With this interpretation the maximal topological order of $W_{1+\infty}$ minimal models would be characterized by non-Abelian, fractionally charged excitations that behave exactly as SU($m$) quarks. The topologically ordered medium corresponds to the asymptotically free regime in which these quarks are liberated; when topological order is destroyed the quarks are confined within integrally charged electrons. 

Having derived the quantum dimensions of excitations we are now ready to compute the topological entanglement entropy \cite{kit1} of the $W_{1+\infty}$ minimal models and to compare it with the corresponding Abelian Chern-Simons models. The topological entanglement entropy $\gamma $ in (\ref{a}) of a topological order is defined as:
\begin{equation}
\gamma =  \log \sqrt{\sum_a{ \cal D}_a^2} \ ,
\label{aa}
\end{equation}
where the sum runs over all superselection sectors. 

Before applying this formula blindly to the $W_{1+\infty}$ minimal models, however, one must remember that there are actually two definitions of the quantum dimension. The first is (\ref{c}). The second is as follows. Imagine fusing $n$ degenerate representations $M$ of $W_{1+\infty}$ and computing the multiplicity of M itself in the resulting decomposition. If this multiplicity is given by $d^n$ in the limit $n\to \infty$, then the quantum dimension ${\cal D}_M = d$. It is this second definition of quantum dimension that enters primarily in the computation of the topological entanglement entropy \cite{kit1}. In the case of rational conformal field theories these two definitions always coincide: the $W_{1+\infty}$ minimal models, however, are not rational. 

Nonetheless, it is easy to convince oneself that also in this case the two definitions of the quantum dimension coincide. The fusion rules of $W_{1+\infty}$ degenerate representations are isomorphic to the decomposition of SU($m$) tensor representations and these fusion rules are the only ingredient needed in the computation of the quantum dimension according to the second definition. Therefore, the quantum dimensions so computed cannot be different from the classical dimensions of SU(m) tensor representations, exactly the result obtained by the computation according to the first definition. 

The final ingredient needed to compute the topological entanglement entropies of the two models are the superselection sectors. Indeed, in both models the  lattice of excitations is infinite, while the concept of entanglement entropy only makes sense for a finite number of superselection sectors. As explained beautifully in \cite{ffn}, this is obtained by maximally enlarging the symmetry algebra to contain also the vertex operators for the creation and annihilation of electrons. In the Abelian Chern-Simons model this corresponds to considering the lattice of excitations modulo the $m$ vectors $(p+1,p,\dots, p)$, $(p,p+1,p,\dots, p)$, $\dots$, $(p,\dots, p,p+1)$ representing the electron excitations in the $m$ components. For simplicity's sake we shall focus on the case $c=2$ and $\nu =2/2p+1$. In this case, the remaining non-trivial superselection sectors are given by the lattice points 
$(1,0) (2,0) \dots (2p+1,0)$.
There are $2p+1$ of these excitations, each one with quantum dimension 1 since they are Abelian. This gives the topological entanglement entropy 
\begin{equation}
\gamma_{CS} (c=2, \nu = 2/2p+1) =  {1 \over 2} \log \left(  2p+1 \right) \ .
\label{ac}
\end{equation}

In the $W_{1+\infty}$ minimal models, the original component label of excitations of the Chern-Simons models is turned by the reduction process into an isospin index. Correspondingly, there are not anymore two electron excitations $(p+1,p)$ and $(p,p+1)$ with an Abelian component index but rather only one electron excitation $(p+1,p)$ (the second one is forbidden since $n_2>n_1$) with isospin 1/2. The two states in the isospin doublet correspond to electrons in the two different components. In this case it is natural to mode out the excitation lattice by the vector $(p+1,p)$ corresponding to the isospin-carrying electron and by the vector $(1,-1)$ corresponding to the elementary neutral excitation with isospin 1.

Moding out by these two vectors gives indeed the same number of superselection sectors as in the Abelian Chern-Simons model. The quantum dimensions, however are now given by ${\cal D}_s=2s+1$
where $s=|n_1-n_2|/2$ is the isospin of the excitation for $s = 0, 1/2$. This gives
\begin{equation}
\gamma_{W_{1+\infty}}  (c=2, \nu = 2/ 2p+1) = {1 \over 2} \log {( 5 p + 4 ) } \ .
\label{ad}
\end{equation}

This is larger than the corresponding quantity in the Abelian Chern-Simons models and, correspondingly, the total entanglement entropy is lower. This is because the topological order embodied in the $W_{1+\infty}$ minimal models is more robust than its Abelian Chern-Simons counterpart. We believe that it is this type of maximal topological order that is realized in the composite fractional quantum Hall fluids and we propose the above topological entanglement entropy as a possible experimental measure to confirm this. The confirmation of this type of topological order would suggest non-Abelian anyons in the $\nu=2/5$ quantum Hall states as primary candidates for topological quantum computation.

\end{document}